\documentclass[9t,twocolumn,twoside]{opticajnl}

\journal{opticajournal} 

\setboolean{shortarticle}{true}

\usepackage{lineno}
\usepackage{hyperref}

\title{Coherent amplitude modulation of continuous-wave light in cesium vapor}

\author[1,2]{X. Zhang}
\author[3]{ J. B. Kim}
\author[4,*]{ D. Antypas}

\affil[1]{Johannes Gutenberg-Universit{\"a}t Mainz, GSI Helmholtzzentrum f{\"u}r Schwerionenforschung, 55128 Mainz, Germany}
\affil[2]{School of Control Science and Engineering, Dalian University of Technology, Dalian 116024, China}

\affil[3]{Department of Physics Education, Korea National University of Education, Cheongju 28173, Korea}

\affil[4]{Department of Physics, University of Crete, 70013 Heraklion-Crete, Greece}

\affil[*]{dantypas@physics.uoc.gr}

\begin{abstract}
We report on observations of coherent, sustained oscillations in the absorption of  continuous-wave light at 388 nm that excites the $6S_{1/2}\rightarrow 8P_{3/2}$ transition in cesium vapor. The oscillation frequency is close to the spacing of hyperfine levels of the $8P_{3/2}$ level that are excited simultaneously by the 388 nm field. We observe threshold behavior of the oscillation amplitude with pump power, and suggest that the effect is associated  with infrared directional emission due to amplified spontaneous emission from the $8P_{3/2}\rightarrow 8S_{1/2}$ transition, that is assisted by retro-reflections from the cell windows. The effect may be used to probe a lasing process in an atomic vapor, by checking the temporal properties of the pump field transmitted through the vapor.

\end{abstract}

\setboolean{displaycopyright}
{false} 

\begin{document}

\maketitle

Since thermal noise of the laser cavity limits the performance of today's best atomic clocks, an alternative approach based on a superradiant laser, in which the optical clock transition in alkaline earth atoms acts as the gain medium, has been proposed \cite{MeiserPRL2009}. In this case the gain bandwidth is much less than the cavity linewidth as with masers in the microwave domain.  It has been shown that the ultimate linewidth of optical radiation generated in this so-called bad-cavity regime can be smaller than the value given by the Schawlow-Townes expression due to the frequency pooling effect \cite{MeiserPRL2009, KuppensPRL1994}. The laser would be rather insensitive to the mirror fluctuations.
In addition to state-of-the-art optical atomic clocks with record stability and reproducibility, the bad-cavity approach could be useful for creating stable optical local oscillators in the telecommunications bandwidth \cite{XuCPL2015} with potential for miniaturization. 

Furthermore, there are hints that lasing of elongated population-inverted atomic medium due to the process of amplified spontaneous emission (ASE)\cite{PetersJPA1971, CaspersonJAP1977} may also have narrow linewidths without any cavity \cite{AkulshinAPB2013, BrekkeOL2015}. ASE-based lasing requires more focused research, especially the influence of weak optical feedback, which can play a significant role in determining the spectral and intensity properties of the generated light.
%


Here we discuss an effect arising due to ASE-based lasing that is assisted by weak feedback. 
Specifically, we report on observations of highly coherent, sustained oscillations in the intensity of continuous-wave (cw), narrow-band light transmitted through cesium (Cs) vapor and tuned in frequency to excite an electric-dipole atomic transition. As we will see, the effect may be linked to generated infrared emission at $6.8\,\mu$m due to an ASE process in the vapor. This emission is amplified due to weak optical feedback from the vapor cell windows. Temporal properties of the transmission of a cw field exciting an atomic vapor have been studied in several contexts (see, for example, \cite{CamparoJOSAB1998,PAPOYANOptComm2021}), as well as temporal dynamics of fields generated through ASE and four-wave-mixing-(FWM) processes \cite{AkulshinJOSAB2020, AkulshinOpticsLetters2021}. The observations presented here, however, have not been (to our knowledge) reported previously.

\begin{figure}[ht!]
\centering\includegraphics[width=8cm]{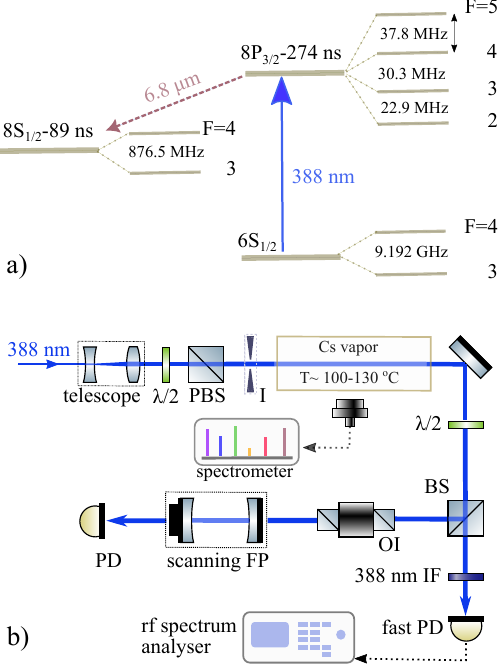}
\caption{a) Partial energy level diagram of Cs. Hyperfine-level spacings computed from \cite{AllegriniJPCRD2022}. b) Experimental apparatus. $\lambda/2$: half-wave plate; PBS: polarizing beamsplitter; I: Iris; OI: optical isolator; FP: Fabry-Pérot  spectrum analyzer; PD: photodetector; IF: interference filter.}

 
\label{fig:Fig1ab}
\end{figure}

The observed effect is on the Cs $6S_{1/2}\rightarrow 8P_{3/2}$ transition at 388 nm (Fig.\,\ref{fig:Fig1ab}a). The oscillation frequency is found to be significantly greater than the natural linewidth of any energy levels below the excited $8P$ level. This frequency is close to the frequency spacing between two hyperfine levels of the  $8P_{3/2}$ level, while the respective spectral feature is rather narrow, typically having several kHz in width, i.e. much lower than  the $\approx$580 kHz natural linewidth of the  $6S_{1/2}\rightarrow 8P_{3/2}$ transition. Moreover, it is seen that the oscillation amplitude exhibits threshold behavior with the power of the UV excitation field, analogously to the emergence of directional emission due to an ASE process, as studied in the context of mirrorless lasing (see, for example, \cite{SharmaAPL1981,SellOL2014,AkulshinOL2014,AkulshinOL2018, AntypasOL2019,SebbagOL2019}). Several ASE processes have been observed in the range $0.76-3.2\,\mu$m in Cs due to pumping at 388 nm \cite{AntypasOL2019}.

Our experiments are carried out with an apparatus whose schematic is shown in Fig.\ref{fig:Fig1ab}b. Light from a frequency-doubled Ti:Sapphire laser (M Squared Sols:TiS \& ECD-X) with  specified linewidth for the fundamental frequency of 50\,kHz is tuned to excite the 388 nm transition in a Cs vapor cell, whose temperature is maintained (to within 1\,$^o$C) in the range 100-135\,$^o$C. The laser beam is linearly polarized and unless stated otherwise below, it is collimated with a $\approx 2.5$\,mm diameter. Its power can be varied in the range $0-130$\,mW. The cell with nearly parallel windows is 72 mm long with a 25 mm diameter and is made of borosilicate glass. Part of the light transmitted through the cell is sent to a scanning Fabry-Pérot (FP) spectrum analyzer, while the remaining part is measured with a fast photodetector (PD) (150 MHz bandwidth), whose output is monitored in a rf spectrum analyzer (Keysight N9320B) or an oscilloscope. Fluorescence from atoms is measured from the side of the cell with a photodetector, or with a fiber that is used to collect light for analysis with a spectrometer that operates in the range 700-900 nm. A 390 nm interference filter with a 10 nm bandwidth ensures that only the pump light is measured with the photodetector. 
The laser frequency is monitored with a wavemeter (High Finesse WS8-2). 

\begin{figure}[ht!]
\centering\includegraphics[width=9cm]{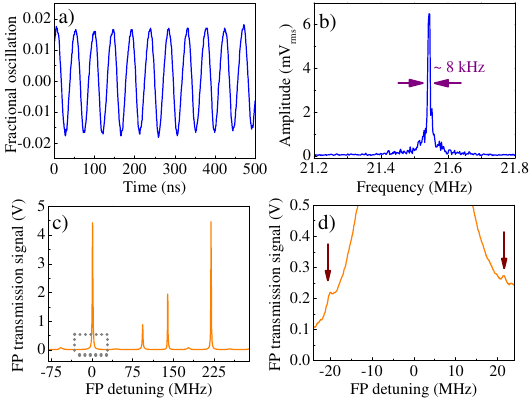}
\caption{Oscillations in the power of  388 nm light transmitted through the Cs vapor cell. a) Single time trace of measured light (no averaging done). b) Amplitude spectrum of the 388 nm light, measured with a rf spectrum  analyser with a set 1 kHz resolution bandwidth. A narrow spectral feature appears at $\approx$21.55 MHz. c) Spectrum of a scanning FP used measure the 388 nm light. The peaks at about 90 and 140 MHz are higher order modes of the cavity. d) FP spectrum shown around the base of the fundamental resonator fringe, indicated by the dotted rectangle in c). The two arrows show sidebands about the main fringe.}
\label{fig:Fig2}
\end{figure}

We illustrate the observed effect in  Fig.\,\ref{fig:Fig2}a. When the laser frequency $\nu_L$ is tuned in the vicinity of the Doppler-broadened $6S_{1/2}\rightarrow 8P_{3/2}$ resonance (of $\approx$1\,GHz linewidth), the transmitted radiation through the Cs cell obtains an oscillating part, with an amplitude that (depending on 388 nm power) can be as large as few \% of  average intensity level. In the example of Fig.\,\ref{fig:Fig2}a, $\nu_L$ is tuned to excite atoms from the $F=3$ hyperfine level of the ground $6S_{1/2}$ level, resulting in pumping to the $F=2,3$ and $4$ levels of the excited $8P_{3/2}$ level, where $F$ denotes the total angular momentum of a level. For the $\approx$130$^o$C vapor cell temperature where the effect is more pronounced, oscillations appear for $\nu_L$ detuning of $\approx$800\,MHz below the resonance center (where $\approx$30\% of the 388 nm beam is absorbed by atoms), and within a $\approx$40\,MHz window. Oscillations vanish completely outside this frequency window. (We note that oscillations were observed for cell temperatures as low as 95\,$^o$C; at lower temperature, the required 388 nm detuning is lower, for example, $\approx 120$\,MHz at 95\,$^o$C .) The corresponding spectrum of the fast PD signal shows a peak that (depending on experimental conditions) lies $1-2$\,MHz below the  $8P_{3/2}$ level's $F=2$ to $F=3$ hyperfine level spacing  $\Delta_{2-3}=22.9$\,MHz (Fig.\,\ref{fig:Fig2}b). Remarkably, the peak exhibits a rather narrow width, typically observed to be several kHz.

Oscillations also appear when tuning $\nu_L$ to pump atoms out of the ground $F=4$ level, resulting in excitations to the excited $F=3,4$ and 5 levels. In this case, the oscillations have frequency $\approx35$\,MHz, i.e. close to the $8P_{3/2}$ level's F=4 to F=5 spacing ($\Delta_{4-5}=37.8$\,MHz).

Looking at the spectrum of the UV light transmitted through the cell with the FP analyzer, we see two small side-fringes about a main interferometer fringe (Fig.\,\ref{fig:Fig2}c and Fig.\,\ref{fig:Fig2}d), at a distance from the main fringe equal to the frequency of the observed oscillations. This set of side peaks is expected in case of amplitude modulation for the transmitted light, where two sidebands appear in the spectrum.

\begin{figure}[ht!]
\centering\includegraphics[width=9cm]{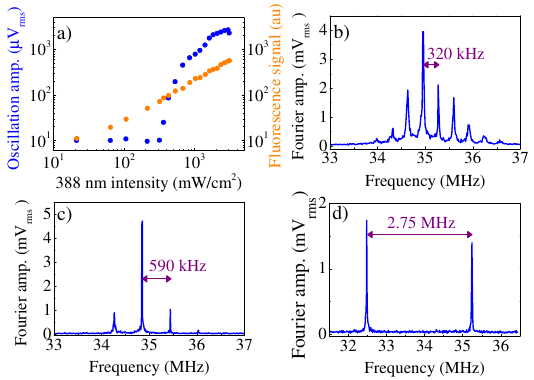}
\caption{a) Oscillation amplitude and IR fluorescence level with pump intensity. Fluorescence is recorded in the $700-900$\,nm range with the spectrometer. b), c) and d): Amplitude spectra of transmitted light 
 showing multiple oscillation frequencies. }
\label{fig:Fig3}
\end{figure}

The oscillations do not grow proportionally in amplitude with the power of the UV field. We find a threshold for their emergence, unlike fluorescence emission that rises linearly with power  (Fig.\,\ref{fig:Fig3}a). This hints to the effect possibly being associated with an ASE process in the vapor, that is known to exhibit threshold behavior. 

We find that the effect occurs when the 388 nm beam is retro-reflected from the exit cell window, so that it overlaps the input beam. This alignment is done by orienting the cell to obtain visual overlap of the respective fluorescence columns within the cell. No oscillations are seen under any conditions if retro-reflection is spoiled, or when an alternative Cs cell with tilted windows is tested. 

In addition, the effect is observed when light reflection occurs exclusively from the inner surface of the exit window. No oscillations are seen under any conditions when retro-reflection is done externally with a partial (50\%) reflector that is used to send part of the transmitted UV beam back into the cell. Since an external retro-reflection is found to produce no oscillations but an internal does, it is conceivable that oscillations are crucially linked to a retro-reflected field other than the UV field. A necessary assumption for this is that the assumed field is fully absorbed at the exit window, so that only an internal retro-reflection may effect the oscillations.  

Considering the above phenomenology and assumption, we suggest that directional emission at 6.8\,$\mu$m that arises due to ASE in the $8P_{3/2}\rightarrow 8S_{1/2}$ channel (Fig.\,\ref{fig:Fig1ab}a) and is assisted in efficiency by retro-reflections at the cell windows, is linked to the observed oscillations. Such an emission is expected in the presence of population inversion between the $8P_{3/2}$ and $8S_{1/2}$ levels in a pencil-shaped region, determined by the pump beam. Directional emission at 6.8\,$\mu$m would be fully absorbed by the borosilicate-glass window of the employed Thorlabs cell (with known transmission curve), unlike all other strong ASE processes (occurring in range 0.76-3.2 $\mu$m \cite{AntypasOL2019}) in Cs that arise due to pumping of atoms to the $8P_{3/2}$ level. Since these atoms have different velocities along the UV pump beam, they are excited to different sublevels of the $8P_{3/2}$ level. 
As population inversion on the $8P_{3/2}-8S_{1/2}$  transition can be prepared within several groups of atoms with different velocities, the process of ASE can generate several directional fields at the 6.8\,$\mu$m within the $8P_{3/2}(F=3,4,5)\rightarrow 8S_{1/2}(F=3,4)$ manifold. The Doppler shift at 6.8\,$\mu$m for the groups of atoms pumped by the $\approx800$\,MHz red-detuned UV field is only about 46\,MHz (and, respectively about 7\,MHz for oscillations observed at low cell temperature, for which the 388 nm detuning is 120\,MHz).  Thus, it is conceivable that the forward and reflected IR fields of given frequency both interact with the different groups of atoms pumped by the UV field. Presumably, it is the generation of these multiple IR fields  at $6.8$\,$\mu$m
that results in coherent modulation in the atomic populations of the $6S_{1/2}$ and $8P_{3/2}$ levels. 

Insight into the process creating this modulation may be gained by consideration of the observed oscillation frequency for the transmitted UV field intensity.  This frequency is generally slightly below \mbox{($\approx$\, $0.5-3$\,MHz)} the spacing between two adjacent $8P_{3/2}$ hyperfine levels, where one of the two can only decay to a single $8S_{1/2}$ hyperfine level, while the other is allowed to decay to the latter level. For instance, when exciting the $6S_{1/2}\,F=3\rightarrow 8P_{3/2}\,F=2,3,4$ transitions, oscillation is observed at frequency slightly smaller than the  $8P_{3/2}$ level's $F=2$ to $F=3$ spacing  $\Delta_{2-3}=22.9$\,MHz (the $F=2$ only decays to the $8S_{1/2}$\,$F=3$ level and the $F=3$ may as well decay to the latter). Similarly, with the UV field tuned to excite the $6S_{1/2}\,F=4\rightarrow 8P_{3/2}\,F=3,4,5$ transitions, the oscillation frequency is slightly below the $8P_{3/2}$ level's $F=4$ to $F=5$ spacing $\Delta_{4-5}=37.8$\,MHz. A pair of ASE-based $6.8\,\mu$m fields with frequency difference close to the hyperfine level spacing  $\Delta_{2-3}$ or $\Delta_{4-5}$ can be thus associated with the modulation in the populations of the ground and excited levels of the 388 nm transition. Based on this, one may predict such modulations to be detectable in other systems. One such case is the $^{87}$Rb $5S_{1/2}\rightarrow 6P_{3/2}$ transition at 420 nm, where the excited level hyperfine spacing is in range $24-67$\,MHz, and the strong decay   $6P_{3/2}\rightarrow 6S_{1/2}$ yielding ASE \cite{MoranOC2016} is in the IR (2.73\,$\mu$m), so that the ASE-based fields in the vapor may interact with several velocity groups pumped by the 420 nm field.

It is possible that the $6.8\,\mu$m emission is part of FWM processes in the parametric loops $6S_{1/2}\rightarrow 8P_{3/2}\rightarrow 8S_{1/2}\rightarrow
7P_{3/2,1/2}\rightarrow
6S_{1/2}$. These processes result in blue light generation at 456 and 459 nm \cite{AntypasOL2019}. Such blue light was not observed under any conditions when using a diffraction grating to analyse the spectral content of light emerging from the vapor cell in the forward direction.

In passing, we note that oscillations were not observed when exciting atoms in the  $6S_{1/2}\rightarrow 8P_{1/2}$ transition at 389 nm.  The hyperfine-level spacing between the $F=3$ and $F=4$ levels of the $8P_{1/2}$ level (171.8 MHz) is significantly larger than the  $22.9-37.8$\,MHz spacings in the manifold of the  $8P_{3/2}$ level. 
 
There are additional observations related to the rapid modulation in the populations of $6S$ and $8P$ levels. As we found, oscillations do not always appear at a single frequency. Indeed, it is possible via certain adjustment of the 388 nm laser frequency, power, beam profile, and  alignment (i,e. the angle between the cell and the laser beam), to obtain spectra showing multiple sidebands. We illustrate such a spectrum in Fig.\,\ref{fig:Fig3}b. Atoms are excited out of the $F=4$ ground level, where multiple-sideband spectra are more efficiently produced. The cell temperature is $T=131$\,$^o$C, the laser detuning from the Doppler-broadened resonance is $\approx$\,920 MHz below its center,
and the optical power is $\approx$54\,mW. Several sidebands are seen in the spectrum of Fig.\,\ref{fig:Fig3}b, spaced by $\approx$320\,kHz. Furthermore, repeating this experiment with a divergent optical beam entering the cell (divergence half-angle\,$\approx$\,3\,mrad, beam diameter at the cell center $\approx 5$\,mm), we found a pattern of sidebands with larger spacing, shown in  Fig.\,\ref{fig:Fig3}c. We additionally looked for a sideband pattern while adjusting with an iris (Fig.\,\ref{fig:Fig1ab}b) the diameter of the divergent beam entering the vapor cell.  Starting with an optical power of 120 mW, and shrinking the beam diameter with the iris, we obtained a dual-peak spectrum shown in Fig.\,\ref{fig:Fig3}d. The two peaks are distant by several MHz. A recorded example of the appearance of sidebands as beam diameter is varied, is provided in the Visualization 1. 

The observed sideband emergence may also be due to reflections from the two cell windows, whose inner faces create a weak cavity for directional emission at 6.8\,$\mu$m. In this case, the observed spacings in the spectra shown in Fig.\,\ref{fig:Fig3} could  be linked to the frequency spacing of transverse electromagnetic modes that are dominant in this cavity. The latter spacings would be determined by various aspects such as beam alignment, pump beam divergence or diameter. 

Insight into the mechanism generating the observed effect could be gained by directly detecting the 6.8\,$\mu$m field generated in the vapor. This requires use of a cell with windows allowing some transmission of the 6.8\,$\mu$m field. A detailed simulation of the cooperative effects arising in the population-inverted atoms in the presence of a weak cavity for the 6.8\,$\mu$m radiation, may also help identify the origin of our observations. If established that the pump beam oscillations are linked to the emergence of directional emission, the observed effect may be used to indirectly probe lasing in atomic vapors by looking at the pump field's temporal properties.
Potentially, the system studied here may be employed to develop a stable optical clock in the bad cavity regime.


\textbf{Funding.} J.B. Kim is grateful for the support of the National Research Foundation of Korea (NRF-2021R1F1A1060385). This work was supported by the European Research Council (ERC) under the European Union Horizon 2020 research and innovation program (project YbFUN, grant agreement No 947696).

\textbf{Acknowledgment.} Authors acknowledge A. Akulshin, D.S. Elliott, and D. Budker for insightful discussions.

\textbf{Disclosures.} 
The authors declare no conflicts of interest.

\textbf{Data availability.} Data underlying the results presented in this
paper are not publicly available at this time but may be obtained
from the authors upon reasonable request.

\smallskip

\section{References}

\bibliography{sample}

\begin{thebibliography}{10}
\newcommand{\enquote}[1]{``#1''}

\bibitem{MeiserPRL2009}
D.~Meiser, J.~Ye, D.~R. Carlson, and M.~J. Holland,
  {\protect\JournalTitle{Phys. Rev. Lett.}} \textbf{102}, 163601 (2009).

\bibitem{KuppensPRL1994}
S.~J.~M. Kuppens, M.~P. van Exter, and J.~P. Woerdman,
  {\protect\JournalTitle{Phys. Rev. Lett.}} \textbf{72}, 3815 (1994).

\bibitem{XuCPL2015}
Z.-C. Xu, D.~Pan, W.~Zhuang, and J.-B. Chen, {\protect\JournalTitle{Chinese
  Physics Letters}} \textbf{32}, 093201 (2015).

\bibitem{PetersJPA1971}
G.~I. Peters and L.~Allen, {\protect\JournalTitle{Journal of Physics A: General
  Physics}} \textbf{4}, 238 (1971).

\bibitem{CaspersonJAP1977}
L.~W. Casperson, {\protect\JournalTitle{J. of Appl. Phys.}} \textbf{1}, 256
  (1977).

\bibitem{AkulshinAPB2013}
A.~Akulshin, C.~Perrella, G.-W. Truong, A.~Luiten, D.~Budker, and R.~Mclean,
  {\protect\JournalTitle{Applied Physics B}} \textbf{117}, 203 (2013).

\bibitem{BrekkeOL2015}
E.~Brekke and E.~Herman, {\protect\JournalTitle{Opt. Lett.}} \textbf{40}, 5674
  (2015).

\bibitem{CamparoJOSAB1998}
J.~C. Camparo, {\protect\JournalTitle{J. Opt. Soc. Am. B}} \textbf{15}, 1177
  (1998).

\bibitem{PAPOYANOptComm2021}
A.~Papoyan and S.~Shmavonyan, {\protect\JournalTitle{Optics Communications}}
  \textbf{482}, 126561 (2021).

\bibitem{AkulshinJOSAB2020}
A.~M. Akulshin, N.~Rahaman, S.~A. Suslov, D.~Budker, and R.~J. McLean,
  {\protect\JournalTitle{J. Opt. Soc. Am. B}} \textbf{37}, 2430 (2020).

\bibitem{AkulshinOpticsLetters2021}
A.~Akulshin, F.~P. Bustos, and D.~Budker, {\protect\JournalTitle{Opt. Lett.}}
  \textbf{46}, 2131 (2021).

\bibitem{SharmaAPL1981}
A.~Sharma, N.~D. Bhaskar, Y.~Q. Lu, and W.~Happer,
  {\protect\JournalTitle{Applied Physics Letters}} \textbf{39}, 209 (1981).

\bibitem{SellOL2014}
J.~F. Sell, M.~A. Gearba, B.~D. DePaola, and R.~J. Knize,
  {\protect\JournalTitle{Opt. Lett.}} \textbf{39}, 528 (2014).

\bibitem{AkulshinOL2014}
A.~Akulshin, D.~Budker, and R.~McLean, {\protect\JournalTitle{Opt. Lett.}}
  \textbf{39}, 845 (2014).

\bibitem{AkulshinOL2018}
A.~M. Akulshin, F.~P. Bustos, and D.~Budker, {\protect\JournalTitle{Opt.
  Lett.}} \textbf{43}, 5279 (2018).

\bibitem{AntypasOL2019}
D.~Antypas, O.~Tretiak, D.~Budker, and A.~Akulshin, {\protect\JournalTitle{Opt.
  Lett.}} \textbf{44}, 3657 (2019).

\bibitem{SebbagOL2019}
Y.~Sebbag, Y.~Barash, and U.~Levy, {\protect\JournalTitle{Opt. Lett.}}
  \textbf{44}, 971 (2019).

\bibitem{AllegriniJPCRD2022}
M.~Allegrini, E.~Arimondo, and L.~A. Orozco, {\protect\JournalTitle{Journal of
  Physical and Chemical Reference Data}} \textbf{51}, 043102 (2022).

\bibitem{MoranOC2016}
P.~J. Moran, R.~M. Richards, C.~A. Rice, and G.~P. Perram,
  {\protect\JournalTitle{Optics Communications}} \textbf{374}, 51 (2016).

\end{thebibliography}

\bibliographyfullrefs{sample}


\ifthenelse{\equal{\journalref}{aop}}{%
\section*{Author Biographies}
\begingroup
\setlength\intextsep{0pt}
\begin{minipage}[t][6.3cm][t]{1.0\textwidth} 
  \begin{wrapfigure}{L}{0.25\textwidth}
    \includegraphics[width=0.25\textwidth]{john_smith.eps}
  \end{wrapfigure}
  \noindent
  {\bfseries John Smith} received his BSc (Mathematics) in 2000 from The University of Maryland. His research interests include lasers and optics.
\end{minipage}
\begin{minipage}{1.0\textwidth}
  \begin{wrapfigure}{L}{0.25\textwidth}
    \includegraphics[width=0.25\textwidth]{alice_smith.eps}
  \end{wrapfigure}
  \noindent
  {\bfseries Alice Smith} also received her BSc (Mathematics) in 2000 from The University of Maryland. Her research interests also include lasers and optics.
\end{minipage}
\endgroup
}{}

\end{document}